\documentstyle[aps,epsf,prl]{revtex}

\begin{document}
\twocolumn
\title{Spontaneous emission of polaritons from a Bose-Einstein condensate}

\author{Karl-Peter Marzlin and Weiping Zhang}
\address{Department of Physics,
 Macquarie University, Sydney, NSW 2109, Australia
}
\maketitle
\begin{abstract}
We study the spontaneous emission of a partially excited
Bose-Einstein condensate composed of two-level atoms. 
The formation of polaritons induced by the ground-state part of the 
condensate leads to an avoided crossing in the photon 
spectrum. This avoided crossing acts similarly to a photonic band gap
and modifies the spontaneous emission rate.
\end{abstract}
$ $ \\
\pacs{
03.75.Fi, 32.80.-t, 42.70.Qs
}
\narrowtext
It is well-known that the radiation properties of atoms can dramatically
be manipulated by changing the enviroment where the atoms emit photons.
For microcavities it has been demonstrated \cite{cavity} and for
periodic dielectric media predicted \cite{john94,kofman94}
that the supression of spontaneous emission (SE) can be quite high.
In the case of a microcavity this happens because its geometry reduces
the radiation-mode density, whereas in a periodic dielectric medium
SE is supressed due to the formation of photonic band gaps (PBG).

The recent achievement of Bose-Einstein condensation in magnetic traps
\cite{experimente} has provided a new state of matter where all atoms
share a single macroscopic quantum state. Such a state of matter offers
great opportunities to explore and test new phenomena related to macroscopic
quantum coherence. Recently several authors have theoretically studied
the optical properties of a Bose-Einstein condensate (BEC).
For example, in a homogeneous, extended BEC polaritons
(superpositions of excited atoms and photons)
are formed \cite{shlyapnikov91}. In a spatially confined BEC, the
continuous center-of-mass momentum distribution leads to an increase of
SE \cite{javanainen93,cooper95}. In addition, the stimulated emission can
be increased by the Bose enhancement in a BEC \cite{hope97}.
In the case of two BECs, interference effects can be important
\cite{savage97}.

In the present paper we focus on the study on how the macroscopic quantum
coherence of a ground-state homogeneous BEC can modify the radiation
properties of atoms which are partially excited from the BEC by a
coherent light field. In difference to a microcavity and PBG
materials, the manipulation of the radiation properties in a homogeneously
extended BEC is caused by the formation of polaritons due to the existence
of ground-state macroscopic quantum coherence. 
 
{\em The model:}
We consider a BEC composed of two-level 
atoms which is coupled to the electromagnetic field.
The coupling is described by using minimal coupling in
rotating-wave approximation under neglection of the term quadratic in the 
electromagnetic field. 
The interaction Hamiltonian is then given by
\begin{equation} H_{\mbox{{\scriptsize int}}} = 
      \int d^3 k d^3 k^\prime  \zeta_\sigma(\vec{k}) 
      a_\sigma(\vec{k}) \Psi_g(\vec{k}^\prime) \Psi_e^\dagger (\vec{k}+
      \vec{k}^\prime) +\mbox{H.c.} \; ,
\label{intham}
\end{equation}
with $\zeta_\sigma(\vec{k}) := \omega_0 \vec{d}\cdot 
\vec{\varepsilon}_\sigma(\vec{k}) [\hbar/(2(2\pi)^3\varepsilon_0 \omega_k
]^{1/2}$ for a electromagnetic mode with polarization vector
$\vec{\varepsilon}_\sigma(\vec{k})$.
The Heisenberg equations of motion for the photon annihilation operators 
$a_\sigma(\vec{k})$ and the field operators $\Psi_e$
and $\Psi_g$ for excited and ground-state atoms can be derived easily and
are given by
\begin{eqnarray} 
  i \hbar \dot{\Psi}_e(\vec{k}) &=& \left \{ \frac{\hbar^2 \vec{k}^2}{2M}
  +\hbar \omega_0 \right \} \Psi_e(\vec{k}) 
  \nonumber \\ & &
  + \int d^3k^\prime \sum_\sigma
  \zeta_\sigma(\vec{k}^\prime ) a_\sigma(\vec{k}^\prime ) \Psi_g (
  \vec{k}-\vec{k}^\prime ) 
  \label{edgl} \\
  i \hbar \dot{\Psi}_g(\vec{k}) &=&  \frac{\hbar^2 \vec{k}^2}{2M}
  \Psi_g(\vec{k}) 
  \nonumber \\ & &
  + \int d^3k^\prime \sum_\sigma
  \zeta^*_\sigma(\vec{k}^\prime ) a^\dagger_\sigma(\vec{k}^\prime ) 
  \Psi_e (\vec{k}+\vec{k}^\prime ) 
  \label{gdgl} \\
  i \hbar \dot{a}_\sigma(\vec{k}) &=&  \hbar \omega_k
  a_\sigma(\vec{k}) + \zeta_\sigma(\vec{k}) \int d^3k^\prime 
   \Psi_g^\dagger(\vec{k}^\prime ) \Psi_e (\vec{k}+\vec{k}^\prime ) 
\label{adgl}
\end{eqnarray} 
We have neglected the nonlinear interatomic interaction terms as 
they are small compared to the coupling energy $\hbar \sqrt{\omega_0
\nu_g}$, where we have defined $\nu_g := |\vec{d}|^2 \rho_g/(2\hbar 
\varepsilon_0)$ with $\rho_g$ being the density of atoms in the 
ground-state and $\vec{d}$ denoting the atomic dipole moment.

{\em Macroscopic coherent solution}:
First, we consider the physics in the coherent regime where the atoms in
the internal ground state share a single macroscopic wavefunction $\psi_g$
and interact with a classical laser field 
$a_\sigma^{\mbox{{\scriptsize class}}}(\vec{k})$. This interaction
prepares the BEC into a coherent superposition of the internal ground-state
$\psi_g$ and an excited state $\psi_e$. The equations of motion for these
three collective fields are given by Eqs.~(\ref{edgl}) to (\ref{adgl})
if the quantum field operators are replaced by the corresponding
classical coherent fields.

We are interested in finding a particular solution of these equations
corresponding to a BEC coherently coupled to a running laser wave. We thus 
make the ansatz
$\psi_g (\vec{k}) = (2\pi)^{3/2} \sqrt{\rho_g} \delta(\vec{k})\exp[-i\mu t]$
for a homogeneous ground-state BEC of density $\rho_g$,
%
%
$a_\sigma^{\mbox{{\scriptsize class}}}(\vec{k})=$ 
$\exp [-i \omega_L t]$
$\delta(\vec{k}-\vec{k}_L)$
$\delta_{\sigma,\sigma_L}$
$\Omega_L [2(2\pi)^3 \hbar \varepsilon_0 \omega_{k_L}]^{1/2}/(|\vec{d}| 
\omega_0)$
corresponding to a laser beam with frequency $\omega_L$, Rabi frequency
$\Omega_L>0$, polarization $\sigma_L$, and wavevector $\vec{k}_L$
({\em inside} the BEC), and
$\psi_e (\vec{k}) = (2\pi)^{3/2} \sqrt{\rho_e} \delta(\vec{k}-\vec{k}_L)
\exp [-i (\mu+\omega_L) t] $ describing coherent excited atoms of density 
$\rho_e$ and of momentum $\hbar \vec{k}_L$. Inserting these expressions
into the Heisenberg equations of motions leads to a set of algebraical 
conditions which fix the parameters $\mu$,
$\vec{k}_L$, and $\rho_e$ which we assume to be smaller than $\rho_g$. 
If we neglect the kinetic energy the density of excited
atoms is given by $\sqrt{\rho_e} = \sqrt{\rho_g} \Omega_L/(\mu + \Delta)$,
where $\Delta := \omega_L -\omega_0$ is the detuning of the laser beam.
The wavevector is fixed by $\omega_{k_L} = \omega_L/2 + \sqrt{(\omega_L/2)^2- 
\omega_0^2 \nu_g /(\mu+\Delta)}$.
For $\Delta \leq 0$ ($\Delta \geq 0$) the chemical potential is 
given by $\mu = -\Delta/2 \pm\sqrt{(\Delta/2)^2 +\Omega_L^2}$ 
which implies $\mu +\Delta >0$ ($\mu+\Delta<0$), respectively. Note
that for $\Delta>0$ the expression for $\omega_{k_L}$ implies the
additional constraint $\mu+\Delta >4\nu_g$.

{\em Derivation of polariton modes:}
Having found the macroscopic coherent solution we now address 
the question of how quantum fluctuations will affect it.
The corresponding stability analysis will
determine how the SE process occurs as the system evolves away from the
initial coherent solution. To perform this analysis we
linearise Eqs.~(\ref{edgl}) to (\ref{adgl})
around the coherent fields by inserting 
$\Psi_g = \exp[-i \mu t] \{ \psi_g + \delta \Psi_g \}$ and
$\Psi_e = \exp[-i (\mu+\omega) t] \{ \psi_e + \delta \Psi_e \}$ 
as well as
$a_\sigma = \exp[-i \omega t] \{ a_\sigma^{\mbox{{\scriptsize class}}} + 
\delta a_\sigma \}$ 
and retaining only terms linear in $\delta \Psi_i$ and $\delta a_\sigma$,
which describe the quantum fluctuations around the coherent solution.

The resulting linearized equations of motions can also be derived from an
effective Hamiltonian for the quantum fluctuations,
\begin{equation} 
  H_{\mbox{{\scriptsize fluct}}} =   H_{\mbox{{\scriptsize pol}}}
  +   H_{\mbox{{\scriptsize spont}}} \; ,
\end{equation} 
where
\begin{eqnarray} 
   H_{\mbox{{\scriptsize pol}}} &=& \hbar \int d^3k \left \{ 
   \left [ \frac{\hbar 
   \vec{k}^2 }{2M} -\Delta -\mu \right ] \delta \Psi_e^\dagger 
   \delta \Psi_e 
   \right . \nonumber \\ & & 
   + \left [ \frac{\hbar
   \vec{k}^2 }{2M} -\mu \right ] \delta \Psi_g^\dagger \delta \Psi_g
   + \sum_\sigma (\omega_k - \omega_L) \delta a_\sigma^\dagger 
   \delta a_\sigma
   \nonumber \\ & &
   + \Omega_L \left [ \delta\Psi_g(\vec{k}-\vec{k}_L) \delta \Psi_e^\dagger 
   (\vec{k}) + \mbox{H.c.} \right ]
   \nonumber \\ & &  \left .  + \sum_\sigma {\vec{d}\cdot 
   \vec{\varepsilon}_\sigma \over |\vec{d}|}
   \sqrt{ \frac{\nu_g \omega_0^2}{\omega_k} } \left [
   \delta a_\sigma(\vec{k}) \delta\Psi_e^\dagger (\vec{k}) +
   \mbox{H.c.} \right ] \right \}
\end{eqnarray} 
describes the formation of polaritons and conserves the total number of
particles.
In contrast to previous work \cite{shlyapnikov91},
it also includes an interaction with the ground-state fluctuations
$\delta \Psi_g$ mediated through the laser field $\Omega_L$. But
since the interaction energy $\hbar \sqrt{\nu_g \omega_0^2/\omega_k}$,
(which we from now on will approximate by $\hbar \sqrt{\nu_g \omega_0}$)
is much larger than $\hbar \Omega_0$ it will only give small corrections
to the polariton energy. We therefore can neglect it. For the same reason
we will omit the kinetic energy $\hbar^2 \vec{k}^2/(2M)$ of the
particles as well as the chemical potential $\mu$.

The eigenmodes of $H_{\mbox{{\scriptsize pol}}}$ are then easily found.
To derive them it is of advantage to introduce the 
``non-coupled'' polarization vector
$\vec{\varepsilon}_{\mbox{{\scriptsize NC}}} (\vec{p})$ which is
perpendicular to the photon momentum $\vec{p}$ and the atomic dipole 
moment $\vec{d}$. Since the interaction couples the polarization to 
$\vec{d}$ a mode proportional to this vector describes uncoupled photons
which do not contribute to SE. A second, ``coupled'',  polarization
vector $ \vec{\varepsilon}_{\mbox{{\scriptsize C}}} (\vec{p})$ is introduced
by being perpendicular to $\vec{p}$ and $ \vec{
\varepsilon}_{\mbox{{\scriptsize NC}}} (\vec{p})$. 
It is easy to show
that the scalar product $ \vec{\varepsilon}
_{\mbox{{\scriptsize C}}} (\vec{p}) \cdot \vec{d}$ is then given by
$|\vec{d}| \sin \vartheta_{\vec{p}}$, where $\vartheta_{\vec{p}}$
is the angle between  $\vec{p}$ and $\vec{d}$.
We associate with  $ \vec{\varepsilon}_{\mbox{{\scriptsize C}}} (\vec{p})$
the quantum fluctuation operator $\delta a(\vec{k}) := \delta a_{\sigma = C}
(\vec{k})$.

The relevant eigenmodes of $H_{\mbox{{\scriptsize pol}}}$ then
consist of two parts. 
(i) free ground-state atoms $\varphi_{g,\vec{k}} (\vec{p})
=\delta (\vec{p}-\vec{k})$ with momentum $\hbar \vec{k}$ and energy
$E_{g,\vec{k}} \approx 0$, and
(ii) polaritons with frequency spectrum 
\begin{equation} 
  \omega_{\vec{k}, \pm} = -\Delta \pm w_{\vec{k}}\; ,
\end{equation}  
with $w_{\vec{k}} := \sqrt{(\Delta_k/2)^2 + \nu_g \omega_0 \sin^2 
\vartheta_{\vec{k}}}$ and $\Delta_k := \omega_k -\omega_0$. 
The polariton modes are given by
\begin{equation}  
\varphi_{\pm,\vec{k}} (\vec{p}) = 
    \frac{ \delta(\vec{p}-\vec{k}) }{\sqrt{ 2w_{\vec{p}} }  }
    \left (\pm \frac{ \sqrt{\nu_g\omega_0} \sin \vartheta_{\vec{p}}  }{
                      \sqrt{ w_{\vec{p}} \pm \Delta_{\vec{p}}/2}     } ,
           \quad \sqrt{ w_{\vec{p}} \pm \Delta_{\vec{p}}/2 }  \right )  \; ,
\end{equation} 
where the first component refers to   
$\delta\Psi_e(\vec{p})$ and the second one to $\delta a(\vec{p})$.
The polariton spectrum $\omega_{\pm,\vec{p}}$  clearly
exhibits an avoided crossing around $\Delta_{\vec{p}} =0$ of width
$\sqrt{\nu_g \omega_0} \sin \vartheta_{\vec{p}}$ (see Fig.~1). 
It also contains a small gap whose edge is reached 
in the limit $|\vec{p}| \rightarrow 0$ and $\infty$ \cite{shlyapnikov91}.

The second part of the Hamiltonian,
\begin{equation} 
   H_{\mbox{{\scriptsize spont}}} =  \hbar \omega_0 \sqrt{\nu_e}   
   \int {d^3 k \over \sqrt{\omega_k}} \sin \vartheta_k \left \{
   \delta a^\dagger (\vec{k}) \delta\Psi_g^\dagger 
   (\vec{k}_L-\vec{k}) + \mbox{H.c.} \right \}
\end{equation} 
describes the spontaneous decay of excited atoms in the coherent field.
It depends on the density of coherently  excited atoms $\rho_e$ through
$\nu_e := \rho_e |\vec{d}|^2/(2\hbar \varepsilon_0)$. Since it has a
structure similar to the exponent of a squeezing operator we can expect
that the coherent field configuration (corresponding to the vaccum 
plus an external field) is instable.

{\em Spontaneous emission of polaritons:}
Having derived the polariton modes we are in the position to calculate
the SE rate. This is done by expanding the
field operators in terms of the eigenmodes,
\begin{equation} 
  (  \delta\Psi_e(\vec{k}) , \delta a(\vec{k})  )  = 
  \int d^3p \sum_{n=\pm}
  \varphi_{n, \vec{p}} (\vec{k})  b_{n, \vec{p}}\; .
\end{equation} 
The operator $b_{n, \vec{p}}$ annihilates a polariton in the
associated mode. In this base $H_{\mbox{{\scriptsize pol}}}$
can be written as
\begin{equation} 
   H_{\mbox{{\scriptsize pol}}} =  \int d^3p \sum_{n=\pm}
   \hbar \omega_{n, \vec{p}} \; b^\dagger_{n, \vec{p}} b_{n, \vec{p}}
\end{equation} 

The interaction hamiltonian $H_{\mbox{{\scriptsize spont}}}$ which
is responsible for SE can be written as
\begin{eqnarray} 
  H_{\mbox{{\scriptsize spont}}} &=& \hbar \omega_0 \sqrt{\nu_e}
  \int {d^3k \over \sqrt{\omega_k}} \sum_{n=\pm} \sin \vartheta_k
  \sqrt{  \frac{w_{\vec{k}} + n \Delta_{\vec{k}}/2 }{2 w_{\vec{k}}}  }
  \nonumber \\ & & \times
  \left \{ 
  b^\dagger_{n, \vec{k}}\; \delta\Psi_g^\dagger(\vec{k}_L-
  \vec{k}) + \mbox{H.c.} \right \}  \; .
\end{eqnarray} 

To derive the SE rate we have to determine the time
evolution of polaritons around the coherent field configuration.
We assume that initially all atoms and photons are described by
the macroscopic coherent solution,
or in other words, the quantized polariton field
is initially in the vacuum $|0 \rangle$. 
This state then evolves under the action of the polariton Hamiltonian. 
To describe this state $|\psi(t)\rangle$ we make the following ansatz,
which corresponds to the one-photon approximation,
\begin{equation} 
  |\psi(t)\rangle \approx \alpha (t) |0 \rangle +
  \int d^3k \sum_{n=\pm} 
  \beta_{n, \vec{k}}(t)
  b^\dagger_{n, \vec{k}}\; \delta\Psi_g^\dagger(\vec{k}_L-  \vec{k}) 
  \;  |0 \rangle  \; .
\end{equation} 
The Schr\"odinger equation $i \hbar | \dot{\psi} \rangle = 
H_{\mbox{{\scriptsize fluct}}}  | \psi\rangle $ then can be solved
by using the Laplace transform $\bar{\alpha}(s) =
\int_0^\infty \exp[-ts] \alpha(t) dt$. In a standard manner
\cite{john94} we then arrive at the equation for the Laplace transform,
\begin{equation} 
  \bar{\alpha}(s) = \alpha(0) \left [ s + I(s) \right ]^{-1}\; .
\label{lapltransf} \end{equation} 
$I(s)$ can be written in the form
\begin{equation} 
  I = \frac{V \nu_e \omega_0^2}{(2\pi)^3} \int d^3k
    \frac{\sin^2 \vartheta_{\vec{k}}}{
    \omega_{\vec{k}} \left ( s-i\omega_L + i \omega_{\vec{k}}  +
    \frac{\nu_g \omega_0 \sin^2 \vartheta_{\vec{k}} }{s-i\Delta}
    \right ) } \; ,
\end{equation}
where $V$ denotes the quantization volume. 
This integral agrees with the one found in absence of a BEC 
(which describes SE in free space) by
setting $\nu_g = \Delta =0$. We denote this free space integral by
$I_0 :=I(\nu_g=\Delta=0)$. 
Both integrals are linearly divergent and can be
treated in the way pointed out by Bethe (see, e.g., Ref.~\cite{milonni}),
i.e., we renormalize the integrals by subtracting the free-electron 
contribution,
\begin{equation} 
  I^{\mbox{{\scriptsize Ren}}} := I -
  \frac{V \nu_e \omega_0^2}{(2\pi)^3 i} \int d^3k
    \frac{\sin^2 \vartheta_{\vec{k}}}{
    \omega^2_{\vec{k}} }
\end{equation} 
and $I_0^{\mbox{{\scriptsize Ren}}} = I^{\mbox{{\scriptsize Ren}}}(\nu_g=
\Delta=0)$. These renormalized integrals are only logarithmically
divergent. 

At this point it is customary in the calculation of the free-space
SE to perform the Wigner-Weisskopf approximation by
neglecting the dependence of  $I_0^{\mbox{{\scriptsize Ren}}}(s)$
on $s$. In the presence of a band gap this is inappropriate
due to the strong variation of the mode density around the gap
\cite{john94,kofman94}. Nevertheless, we can perform a generalized 
Wigner-Weisskopf approximation in the following way.
We expect that the typical 
timescale on which SE happens is much larger than the 
optical cycle timescale $1/\omega_0$. From the definition of the inverse
Laplace transform,
\begin{equation} 
  \alpha(t) = \frac{1}{2\pi i} \int_{\epsilon -i \infty}^{\epsilon +i \infty}
  e^{ts} \bar{\alpha}(s) ds \; ,
\end{equation} 
it is clear that the variable $s$ plays more or less the role of
a Fourier-transformed time. We thus expect that only values of $s$ much
smaller than $\omega_0$ contribute significantly to the SE.
This implies that we can neglect $s$ whereever it appears together with
$\omega_0$ or $\omega_L$. Thus, we are allowed to set $s-i\omega_L
\approx -i \omega_L$ in the denominator of $I(s)$ while retaining the
term depending on $s-i \Delta$.

In the case of $I_0^{\mbox{{\scriptsize Ren}}}$ this procedure
immediately reproduces the Wigner-Weisskopf result
$I_0^{\mbox{{\scriptsize Ren}}} \approx N_e \{ (\gamma_0/2) + i 
\Delta_{\mbox{{\scriptsize Lamb}}}^{\mbox{{\scriptsize 2-lev}}} \}$,
where $\gamma_0 := \vec{d}^2 \omega_0^3/(3\pi \hbar \varepsilon_0 c^3)$
denotes the SE rate in free space and $N_e := V \rho_e$
the number of excited atoms in the BEC.
To fix  $\Delta_{\mbox{{\scriptsize Lamb}}}^{\mbox{{\scriptsize 2-lev}}} $
we follow the theory of Bethe (see, e.g., \cite{milonni})
and introduce a cut-off frequency of $m_e c^2/\hbar$ in 
$I_0^{\mbox{{\scriptsize Ren}}}$, where $m_e$ is the electron's mass. 
Calculating the principal value of
the integral then leads to $\Delta_{\mbox{{\scriptsize Lamb}}}^
{\mbox{{\scriptsize 2-lev}}} \approx 2 \gamma_0$.
In contrast to free space the SE rate in a BEC depends on
$\Delta_{\mbox{{\scriptsize Lamb}}}^{\mbox{{\scriptsize 2-lev}}} $
since such a radiative frequency correction shifts the center of the avoided 
crossing (or of a band gap \cite{kofman94}).

It remains to calculate a renormalized expression of the integral
$I^{\mbox{{\scriptsize Ren}}}$ in the presence of a BEC. Fortunately,
this task reduces to integrals proportional to 
$I_0^{\mbox{{\scriptsize Ren}}}$ and a couple of convergent integrals.
Within the generalized Wigner-Weisskopf approximation we find
\begin{eqnarray} 
   I^{\mbox{{\scriptsize Ren}}} &=& \left ( 1 + \frac{4i\nu_g}{5 (s-i\Delta)}
   \right )  I_0^{\mbox{{\scriptsize Ren}}} +
   \frac{ N_e \gamma_0}{5\pi i} 
   \left \{ \frac{47 i\nu_g}{15(s-i \Delta)} 
   \right . \nonumber \\ & & 
   + {8\over 3}  - \frac{(s-i \Delta)}{i\nu_g}
   + \left ( 1-{4i\nu_g\over (s-i \Delta)}\right )
   \nonumber \\ & & \left . \times 
   (1+{(s-i \Delta)\over i \nu_g})^{3/2} \mbox{arcoth}
   (\sqrt{1+{(s-i \Delta)\over i\nu_g}}) \right \}
\label{iren}
\end{eqnarray} 

To perform the inverse Laplace transformation we have to know the analytical
structure of $\bar{\alpha}(s)$ of Eq.~(\ref{lapltransf}). Because of the
complicated structure of Eq.~(\ref{iren}) we eventually have to rely on
numerical methods to determine $\alpha(t)$, but much can be gained by doing a
general analysis first. In general, $\bar{\alpha}(s)$ has several poles and
a branch cut originating from the term including the arccoth in
$I^{\mbox{{\scriptsize Ren}}}$. This cut lies between $s= i\Delta$ and
$s=i(\Delta - \nu_g)$. The inverse Laplace transform $\alpha(t)$ is then
given by the residues of  $\bar{\alpha}(s)$ at the poles $s_i$ plus a 
contribution from the branch cut,
\begin{eqnarray}  
  \alpha(t) &=& a_0 \left \{ \sum_{\mbox{{\scriptsize poles}}} e^{t s_i}
         \mbox{Res}(\bar{\alpha}(s), s\rightarrow s_i)
  \right . \nonumber \\ & &  \left .
         + \lim_{\epsilon\rightarrow 0}
         \int_{i \Delta}^{i(\Delta-\nu_g)} ds\; e^{t s}  [
         \bar{\alpha}(s+\epsilon) - \bar{\alpha}(s-\epsilon) ] \right \} \; ,
\label{atsol}
\end{eqnarray} 
where we have used that the poles seem always to be of first order and
only the straight parts of the integration around the branch cut do contribute.
We see that to each pole corresponds a fraction of excited atoms 
whose decay rate is given by the real part of the pole $s_i$. 
The branch cut corresponds to a
fraction of excited atoms which decays in a non-exponential way. 

The magnitude of the SE modification is determined by the
ratio of $\nu_g$ to $|s-i \Delta|$ which in free space is of the order of
$\gamma_0 N_e$. Since for contemporary
BECs $\nu_g$ is at best in the order of $\gamma_0$ SE is notably changed only
if there are very few excited atoms $N_e$, which implies $|\Delta|
\gg \Omega_L$. As discussed above, the macroscopic coherent
solution implies in this case for $\Delta>0$ the additional
constraint $\Delta >4\nu_g$.
In Fig.~2 the real part of the two dominating poles $s_1, s_2$ is shown
as a function of $\Delta$ for the case $N_e=1$ and  $\nu_g = 2.5 \gamma_0$
(corresponding to an atom density $\rho_g$ of $5\times 10^{14}$ cm$^{-3}$).
For $\Delta>0$ a third pole appears with a very small negative decay rate
($<10^{-3} \gamma_0$).
The occurence of negative decay rates is consistent within the range of
validity of the linearized equations for the quantum fluctuations
and may indicate the formation of an atom-photon bound state
\cite{john90}.
Obviously the change in the SE rate can be quite large for small $|\Delta|$.
According to Eq.~(\ref{atsol}) the fraction of atoms belonging to the
poles can be easily calculated by determining the residue at the poles.
It turns out that the poles whose real part asymptotically approaches
$\gamma_0$ always dominate and that the fraction of atoms belonging to other
poles is significant only for small $|\Delta|$. The same holds for the
fraction corresponding to the branch cut.

If $|\Delta| \gg \gamma_0$ holds the dominant pole $s_1$ can be
calculated by perturbation theory. Its real part (the decay rate)
is then given by
\begin{equation} 
  {1\over 2} \gamma (\Delta)  = - N_e  {\gamma_0 \over 2}   
  \left ( 1- {4 \nu_g \over 5 \Delta} \right )   \; .
\end{equation} 
We see that the SE rate is altered by a factor of $1- 4\nu_g/(5\Delta)$. 
It depends on the sign of $\Delta$ whether SE is increased or decreased.

In conlusion we have shown that for $N_e = O(1)$
the spontaneous emission is suppressed by the presence of
a BEC in the internal ground state. The latter induces an
avoided crossing in the polariton spectrum which acts similarly to an
effective band gap, thereby suppressing spontaneous emission.

{\bf Acknowledgement}: 
 This work has been supported by the Australian Research Council.


\begin{figure}[t]
\epsfxsize=6cm
\hspace{1cm}
\epsffile{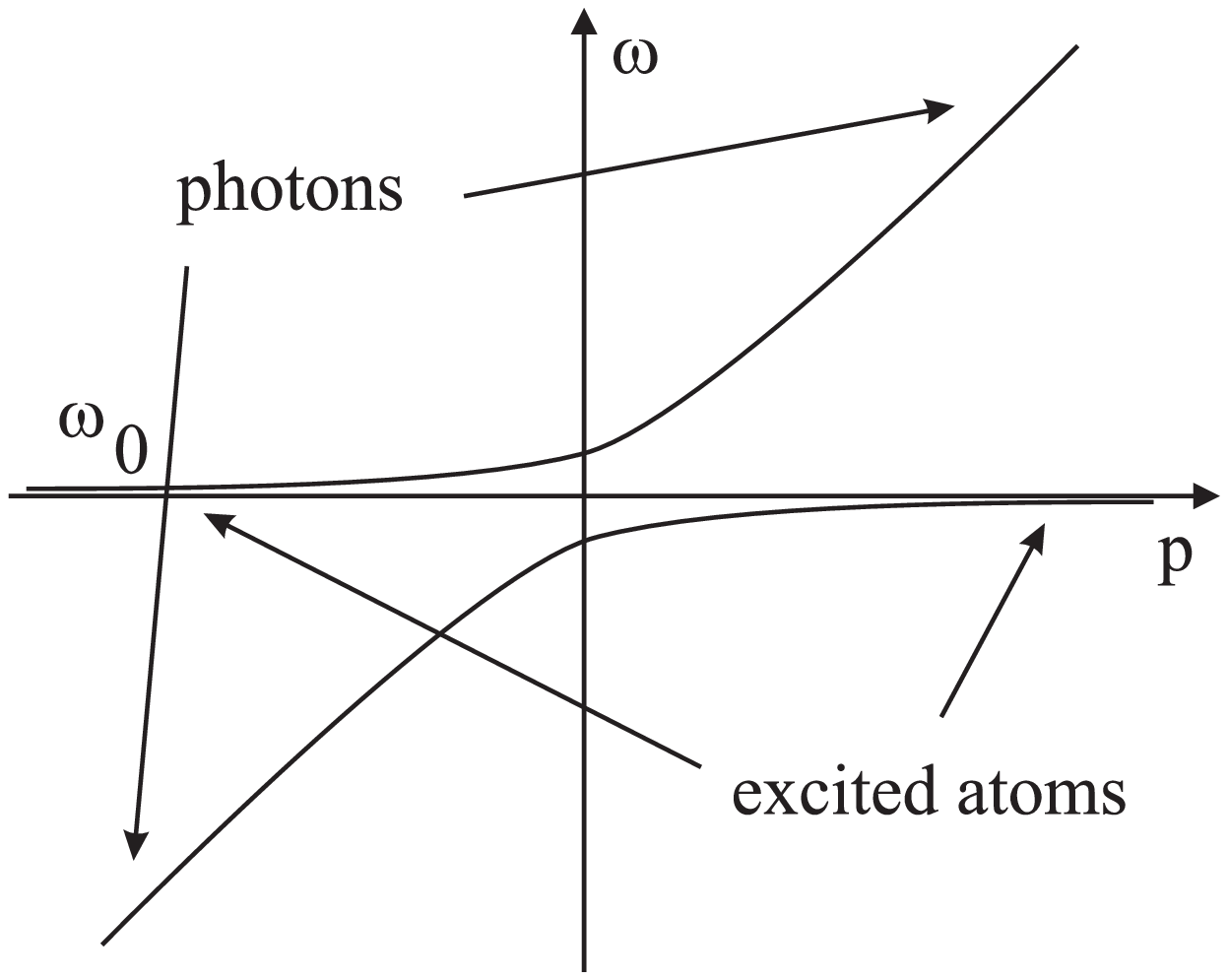}\vspace{-5mm}
\caption{A BEC induces an avoided crossing in the polariton spectrum.
  Far away from the avoided crossing the polaritons describe excited
atoms or photons. Thus, if one focuses on the photons, the avoided crossing
provides an effective band gap.}
\end{figure}

\begin{figure}[t]
\epsfxsize=6cm
\hspace{1cm}
\epsffile{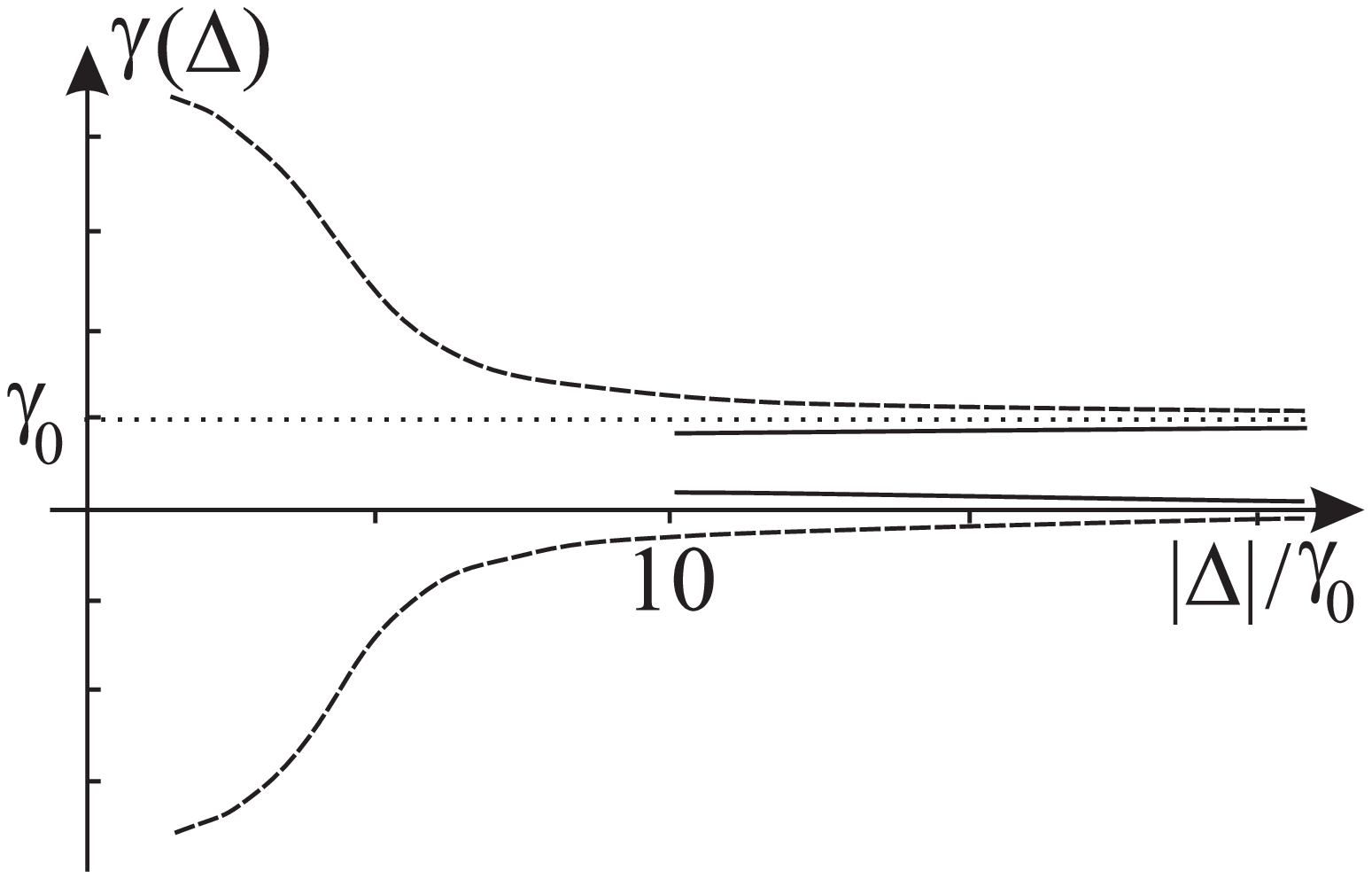}\vspace{-5mm}
\caption{Spontaneous emission rate of a partially excited BEC for 
detuning $\Delta>4 \nu_g$ 
(solid lines) and $\Delta<0$ (dashed lines). The quasi-particles break up
into different fractions with different decay rates. The dominating fraction
is the one whose decay rate asymptotically approaches $\gamma_0$.}
\end{figure}
\end{document}